# AUGMENTING C. ELEGANS MICROSCOPIC DATASET FOR ACCELERATED PATTERN RECOGNITION

**DALI WANG[1], ZHENG LU[1], ZHIRONG BAO[2]**

[1] Department of Electric Engineering and Computer Science, University of Tennessee, TN 37996, USA
[2] Memorial Sloan Kettering Cancer Center, New York, NY 10065, USA
Email: [1]dwang7@utk.edu, [2]zbao@mskcc.org

**Abstract:** The detection of cell shape changes in 3D time-lapse images of complex tissues is an important task. However, it is a challenging and tedious task to establish a comprehensive dataset to improve the performance of deep learning models. In the paper, we present a deep learning approach to augment 3D live images of the *Caenorhabditis elegans* embryo, so that we can further speed up the specific structure pattern recognition. We use an unsupervised training over unlabeled images to generate supplementary datasets for further pattern recognition. Technically, we used Alex-style neural networks in a generative adversarial network framework to generate new datasets that have common features of the *C. elegans* membrane structure. We also made the dataset available for a broad scientific community.

**Keywords**: Adversarial Generative Network, Microscopic Datasets, *C. elegans*, Pattern Recognition.

## 1. INTRODUCTION

Live microscopy and image processing are commonly used for cell dynamic investigation, cellular behavior quantification, and simulation-based hypothesis testing [1-3]. As huge amount of microscope data has been generated during the studies, interactive data analysis become an unprecedented challenge. Although advanced computing technology has been used in microscope data analysis [4], however, these efforts require large datasets with deep domain knowledge. Nowadays, AI-based computer-vision provides a "model-free" approach to solving generic data problems, however, some well-known AI models require massive training datasets. We present a method for augmenting the observation dataset to accelerate the cellular structure image classification using 3D time-lapse datasets directly. We adopt basic concepts within the generative adversarial networks to augment the dataset for speed up the common structure learning. We also work on image noise removal.

## 2. DETAILS EXPERIMENTAL

### 2.1. Materials and Procedures

We use *C. elegans* microscopy images observed from 45 embryos. The raw images (512 x 512 pixel) contain one to three embryos. Raw images are arranged in sets, each with 300 image stacks. These stacks were taken at 1-minute time interval shows the growth of embryos. Each stack is a pseudo 3D image that contains 30 slices showing a different level of the embryo. All the images are captured using the same microscopy setting.

Our first step is to crop raw images into 128x128-pixel images so that each patch contains at most a single embryo and we can use neural networks of moderate size that can fit into a single GPU of an NVIDIA computational platform (see more information at the end of this section). This work is done by an ImageJ macro [5]. For each embryo, we first mark its bounding box, then inside this bounding box, we randomly select image patch of size 128 x 128. Each image patch only contains a part of a single *C. elegans* embryo. For each image patch, we apply a 3-D median filter and adjusting the brightness range to remove the image noise. Examples of a raw image and denoised image are shown in Fig 1. We select the data from a developmental period of 61-minute to 110-minute. For each image stack, we use images between slice 9 and slice 13 as these slices usually have the best imaging quality. For unlabeled dataset, we randomly sampled one image patch of size 128 x 128 at each slice of each image stack. So our dataset contains 45 x 50 x 5 = 11250 image patches.

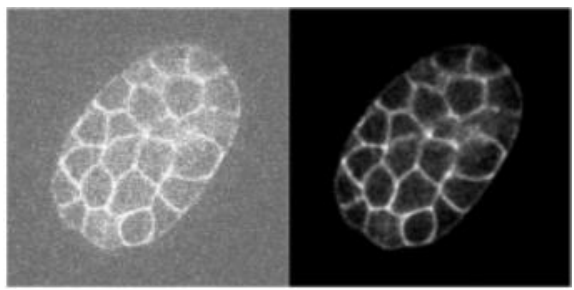

**Fig 1: Microscopy image of C. elegans before and after denoising.**

### 2.2. Computational Platform

We implement our networks with tensorflow 1.7.1, a publicly available deep learning framework. More specifically, the convolutional network for classifying is built upon tensorflow's Estimator API with convolutional network as the customized model function. The generative adversarial network is implemented with tensorflow's TFGAN framework with both generator and discriminator are customized. All experiments are performed on a Nvidia DGX server with four cutting-edge Nvidia Tesla V100 GPUs. Each Tesla V100 is equipped with 640 Tensor Cores and 16 GB memory.

### 2.3. Network Structure

We use an AlexNet-styled convolutional neural network (CNN) [6] to classify the image with particular patterns (i.e., rosettes). CNNs use convolution filters to automatically capture features rather than using hand-engineered features in traditional machine learning algorithms. The network has several convolutional layers (depends on the size of the input image), followed by two fully connected layers. For example, when the input of our network is 128 x 128 grayscale image patches, our network has five convolutional layers. We use 4 x 4 filters for all convolutional layers. The number of filters at the first convolution layer is 32 and doubled at each convolutional layer. Unlike AlexNet, we replace all pooling layers with stride convolutions so that the network can learn its own pooling method. We also place a batch normalization layer after each convolutional layer and the first fully connected layer. Leaky ReLU non-linearities are used as the activations for all layers except the last fully connected layer in the network. Fig 2 showed the details of our convolutional network (for the 128x128 images).

### 2.4 Data augmentation with generative adversarial networks

We adopt several techniques to compensate the potential problems associated small training datasets. We apply dropout during training after the first fully connected layer to eliminate the over-fit problem. We also apply several data augmentation techniques to our dataset including randomly flip the image vertically or horizontally, adjust the brightness and the contrast of the image by a random percentage in a certain range. We use a learning rate of $10^5$ for the training of the network.

It is known then when the size of the dataset is too small for training our AlexNet-style convolutional network, it will result in over-fitted problem. To prove that the network is over-fitted, we first used 198 128x128 images to show the test accuracy of the network in Fig 3 (blue). Then we use a smaller network of half number of features in all convolutional layers and fully connected layers with same settings for the rest of hyper-parameters as shown in Fig 2. It is shown in Fig 3 that the accuracy actually improved with the smaller network. It is important to note that Fig 3 does not imply that the smaller size network is a better choice for the microscope data since the smaller network can capture a limited number of features from large size datasets and the test accuracy of both networks is less than 78\%. Therefore, we decide to develop new methods to improve the network test accuracy using the abundant images without extensive manual annotation.

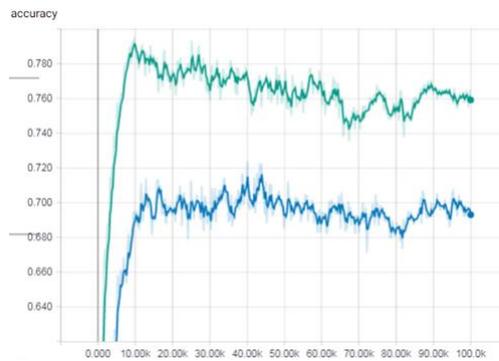

**Fig 3. Overfitting problem. Accuracy improved with smaller network (green)**

In order to further compensate the potential problems associated with small training datasets for image classification and pattern detection, we use Generative adversarial networks (GANs) for data augmentation. GANs is a generative framework that consists of two competing networks: a generator network and a discriminator network. We use a particular form of GAN, called Wasserstein GAN. We use 3 convolutional layer alex-style network structure for both the generator and the discriminator.

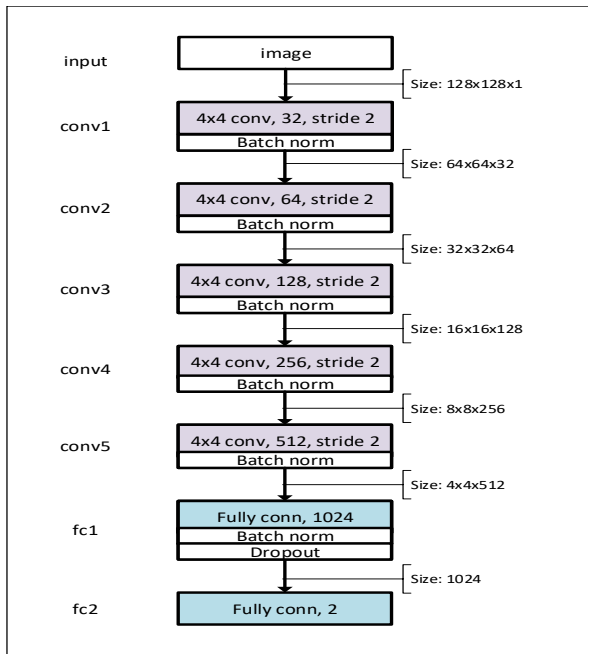

**Fig 2. Structure of the Neural Network.**

The discriminator network has the same network structure as our classifier shown in Fig 2.

Within the GAN framework, the generator produces synthetic data to fool the discriminator while the discriminator network discriminates between real data and synthetic data. The game between the generator $G$ and the discriminator $D$ is the minimax objective:

$$\min_G \max_D E_{x \sim P_{data}}[log D(x)] + E_{\tilde{x} \sim P_g}[log(1 - D(\tilde{x}))]. \quad (1)$$

where $P_{data}$ is the distribution of real data and $P_g$ is the distribution of generated data of $G$ defined by $\tilde{x} = G(z), z \sim P_z$. $z$ is the sample from noise distribution $P_z$, such as the uniform distribution or Gaussian distribution, which is fed to network $G$ as input.

For each update of generator parameters, if the discriminator is trained to optimal, then minimizing the objective function is actually minimizing the Jensen-Shannon (JS) divergence between the real data distribution $P_{data}$ and generated data distribution $P_g$. However, [7] showed that the JS divergence may not be continuous w.r.t generator parameters, so that training of GAN may be hard to converge. To overcome training difficulty, the Wasserstein distance, which is continuous everywhere and differentiable almost everywhere under mild consumption, is proposed to replace JS divergence [8]. Wasserstein distance is also referred to as Earth Mover's Distance (EMD) as it shows the minimum effort to transform one distribution into another.

By using the Kantorovich-Rubinstein duality and a gradient penalty term, the cost function of Wasserstein GAN (WGAN) can be written as:

$$\min_G \max_{D \in C} E_{x \sim P_{data}}[D(x)] - E_{\tilde{x} \sim P_g}[D(\tilde{x})] + P. \quad (2)$$

where $C$ is the set of 1-Lipschitz functions and $P$ is the gradient penalty term.

### 3. RESULTS AND DISCUSSION

We adopt cost function (2) for the WGAN used in our experiments. We show some samples of generated images patches in Fig 4, and then compare them with image patches in real dataset shown in Fig 5. It is shown that the newly generated images in both Fig 4 captured the majority of common features of these 3D images. The Wasserstein losses for both the generator and discriminator of the 128x128 image case are also shown in Fig 6.

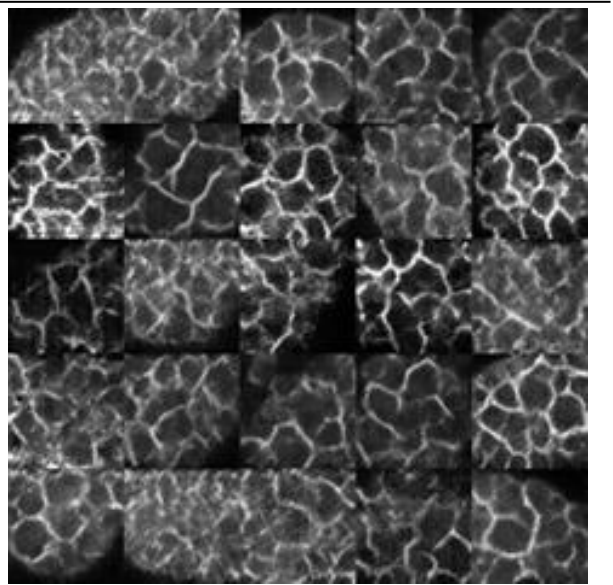

**Fig 4. Generated 128x128 Images.**

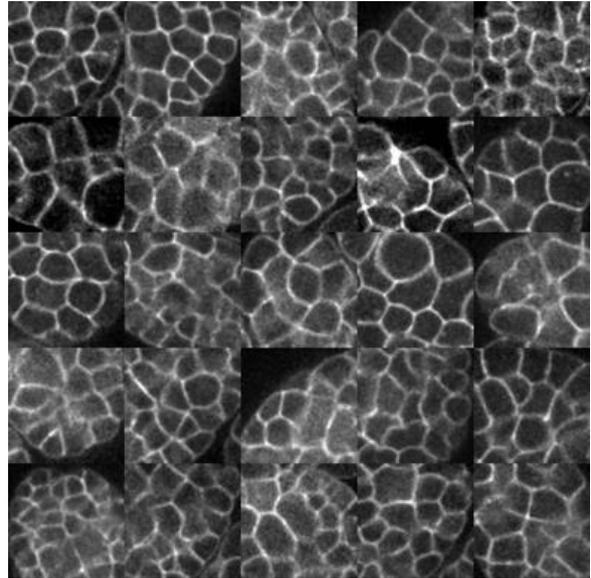

**Fig 5. Generated 128x128 Images.**

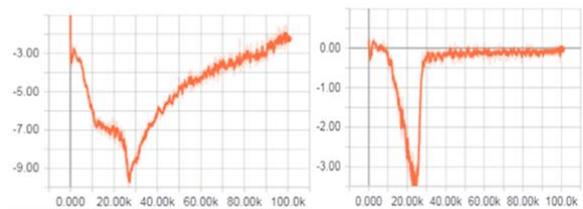

**Fig 6. The Weisserstein loss of the networks (Generator(left), Discriminator (right))**

## CONCLUSIONS

In this paper, we presented a deep learning approach to augment 3D live images of the Caenorhabditis elegans embryo, so that we can further speed up the specific structure pattern recognition. We use an unsupervised training over unlabeled images to generate supplementary datasets for further pattern recognition. Technically, we used Alex-style neural networks in a generative adversarial network framework to generate new datasets that have common features of the C. elegans membrane structure. We also made the dataset available for a broad scientific community.

## ACKNOWLEDGMENTS

This study is supported by an NIH research project grants (R01GM097576). Research in the Bao lab is also supported by an NIH center grant to MSKCC (P30CA008748).

## REFERENCES

1. Bao, Z., Murray, J. I., Boyle, T., Ooi, S. L., Sandel, M. J., and Waterston, R. H. (2006). Automated cell lineage tracing in caenorhabditis elegans. Proceedings of the National Academy of Sciences of the United States of America,103 (8), 2707–2712
2. Wang, Z., Ramsey, B. J., Wang, D., Wong, K., Li, H., Wang, E., and Bao, Z. (2016). An observation-driven agent-based modeling and analysis framework for c. elegans embryogenesis. PloS one , 11 (11), e0166551
3. Wang, Z., Wang, D., Li, C., Xu, Y., Li, H., and Bao, Z. (2018). Deep reinforcement learning of cell movement in the early stage of c. elegans embryogenesis. arXiv, preprint arXiv:1801.04600
4. Jones, T. R., Carpenter, A. E., Lamprecht, M. R., Moffat, J., Silver, S. J., Grenier, J. K., Castoreno, A. B., Eggert, U. S., Root, D. E., Golland, P., and Sabatini,D. M. (2009). Scoring diverse cellular morphologies in image-based screens with iterative feedback and machine learning. Proceedings of the National Academy of Sciences, 106 (6), 1826–1831
5. Schneider, C. A., Rasband, W. S., and Eliceiri, K. W. (2012). Nih image to imagej: 25 years of image analysis. Nature methods,9 (7), 671.
6. Krizhevsky, A., Sutskever, I., and Hinton, G. E. (2012). Imagenet classification with deep convolutional neural networks. In Advances in neural information processing systems, pages 1097–1105
7. Arjovsky, M., Chintala, S., and Bottou, L. (2017). Wasserstein gan. arXiv preprint, arXiv:1701.07875
8. Gulrajani, I., Ahmed, F., Arjovsky, M., Dumoulin, V., and Courville, A. C. (2017). Improved training of wasserstein gans. In Advances in Neural Information Processing Systems, pages 5769–5779

★ ★ ★